\documentclass[epj_mod]{svjour_mod}

\usepackage{graphicx}
\usepackage{amssymb}
\usepackage{amsmath}
\usepackage[bookmarks=true, bookmarksnumbered=true, colorlinks=true, urlcolor=darkblue, citecolor=darkgreen, linkcolor=darkred, breaklinks=true]{hyperref}
\usepackage[all]{hypcap}
\usepackage[usenames,dvipsnames]{color}

\definecolor{darkblue}{rgb}{0,0,0.5}
\definecolor{darkgreen}{rgb}{0.1,0,0.3}
\definecolor{darkred}{rgb}{0.6,0,0}

\def\urltilda{\kern -.15em\lower .7ex\hbox{\~{}}\kern .04em}

\begin{document}

\title{Pippi -- painless parsing, post-processing and plotting of posterior and likelihood samples}
\author{Pat Scott}
\titlerunning{pippi -- parse it, plot it}
\authorrunning{Pat Scott}
\institute{Department of Physics, Imperial College London, London SW7 2AZ, UK\\\email{p.scott@imperial.ac.uk}}
\journalname{Pippi 2.0 manual (arXiv:1206.2245v3)}
\date{This version Nov 17 2016; earlier version published as EPJ Plus 127:138, 2012}

\abstract{
Interpreting samples from likelihood or posterior probability density functions is rarely as straightforward as it seems it should be.  Producing publication-quality graphics of these distributions is often similarly painful.  In this short note I describe \textbf{\textsf{pippi}}, a simple, publicly-available package for parsing and post-processing such samples, as well as generating high-quality PDF graphics of the results.  \textsf{Pippi} is easily and extensively configurable and customisable, both in its options for parsing and post-processing samples, and in the visual aspects of the figures it produces.  I illustrate some of these using an existing supersymmetric global fit, performed in the context of a gamma-ray search for dark matter.  Here I also outline new features introduced in \textsf{pippi 2.0}, including hdf5 support, out of core processing for extremely large datasets, flexible data cuts, per-observable binning, and inline post-processing with arbitrary \textsf{Python} expressions directly from the input \texttt{pip} file.  Pippi can be downloaded and followed at \href{http://github.com/patscott/pippi}{http://github.com/patscott/pippi}.
}

\maketitle

\section{Introduction}
\label{intro}

Many applications in physics and astronomy require sampling from a probability distribution.  Examples include parameter estimation for supersymmetry \cite{Allanach06,Trotta08,Scott09c,Akrami09,Akrami11DD,Akrami11Coverage,Fittino12}, cosmology \cite{CosmoMC,ModeCode1} and cosmic ray propagation \cite{Putze10,GalpBayes}.  A range of sophisticated optimisation and exploration algorithms, and corresponding public codes, exist for doing just this.  These include Markov-chain Monte Carlos (MCMCs; see e.g. \cite{NR3}), nested sampling \cite{Skilling04,MultiNest}, genetic algorithms (e.g. \cite{PIKAIA}) and differential evolution \cite{Storn}.  However, the set of public tools available for analysing samples produced by these algorithms is somewhat smaller, and less developed.  Here I describe \textbf{\textsf{pippi}}, a simple public code for analysing a set of samples from a likelihood or posterior probability density function (PDF).  This updated note serves as an announcement of the public release of \textsf{pippi 2.0}, common documentation of its workings for papers relying on it (e.g.\ \cite{IC22Methods}), and a basic manual for prospective users.

Public codes do exist for this purpose; the best known are \textsf{getdist}, shipped as part of \textsf{CosmoMC} \cite{CosmoMC}, and its various derivatives.  \textsf{Getdist} requires the purchase and installation of \textsf{Matlab}, whereas \textsf{pippi} produces native pdf\LaTeX\ output with \textsf{Python} and the open-source \textsf{Ruby} package \textsf{ctioga2}\footnote{\href{http://ctioga2.rubyforge.org}{http://ctioga2.rubyforge.org/}}.  The resulting plots contain fully embedded \LaTeX\ text and graphics, and are of very high visual quality.  \textsf{Python} rewrites of \textsf{getdist} also exist, and produce similarly high-quality output to \textsf{pippi}.  Apart from the extensive suite of options it offers, \textsf{pippi} differs from those codes in that it is not a translation or rewrite of \textsf{getdist}, and uses interpolation between binned samples rather than a contouring algorithm to produce colour maps; it thus provides a fully independent way to construct distributions from samples.  It has been extensively tested against \textsf{getdist}, and the resulting distributions agree well.

Similar functions are also available as ROOT macros within RooStats \cite{RooStats}. These produce characteristically ugly ROOT figures and require a C++ driver program or implementation within a ROOT session.  \textsf{Barrett} \cite{barrett} and \textsf{Superplot} \cite{superplot} provide additional alternatives.

In the following I briefly describe how \textsf{pippi} works, and give some examples of results produced with it.  I will use the term `chain' to refer to a set of samples produced by an arbitrary sampler, not just an MCMC.

\begin{figure*}[t]
\centering
\begin{minipage}[b]{0.48\textwidth}
a)
\end{minipage}
\hspace{0.01\textwidth}
\begin{minipage}[b]{0.48\textwidth}
b)
\end{minipage}
\begin{minipage}[b]{0.48\textwidth}
\centering
\includegraphics[height=0.85\linewidth]{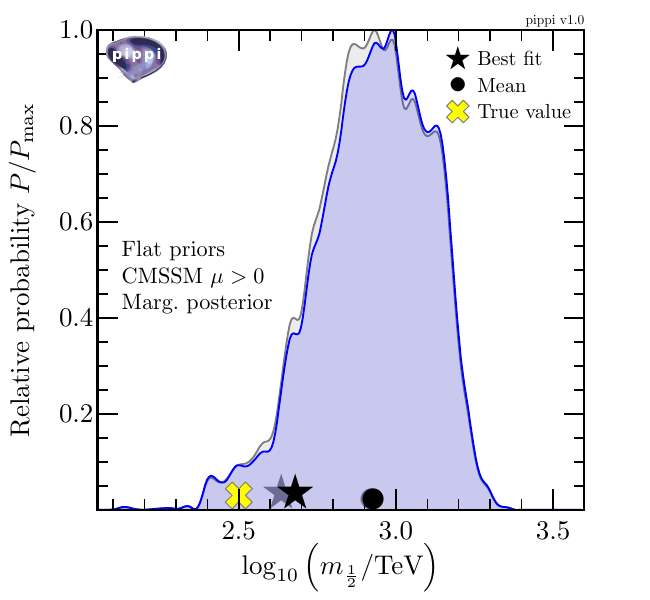}
\end{minipage}
\hspace{0.01\textwidth}
\begin{minipage}[b]{0.48\textwidth}
\centering
\includegraphics[height=0.85\linewidth]{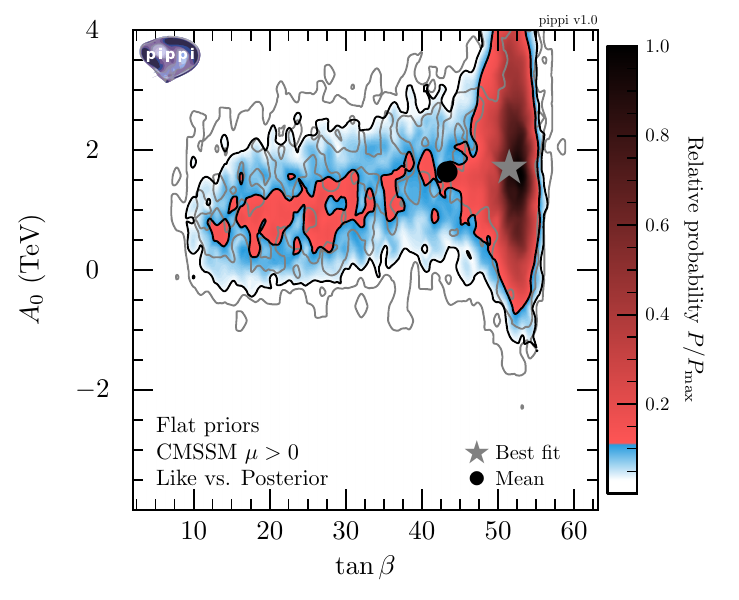}
\end{minipage}
\begin{minipage}[b]{0.48\textwidth}
c)
\end{minipage}
\hspace{0.01\textwidth}
\begin{minipage}[b]{0.48\textwidth}
d)
\end{minipage}
\begin{minipage}[b]{0.48\textwidth}
\centering
\includegraphics[height=0.85\linewidth]{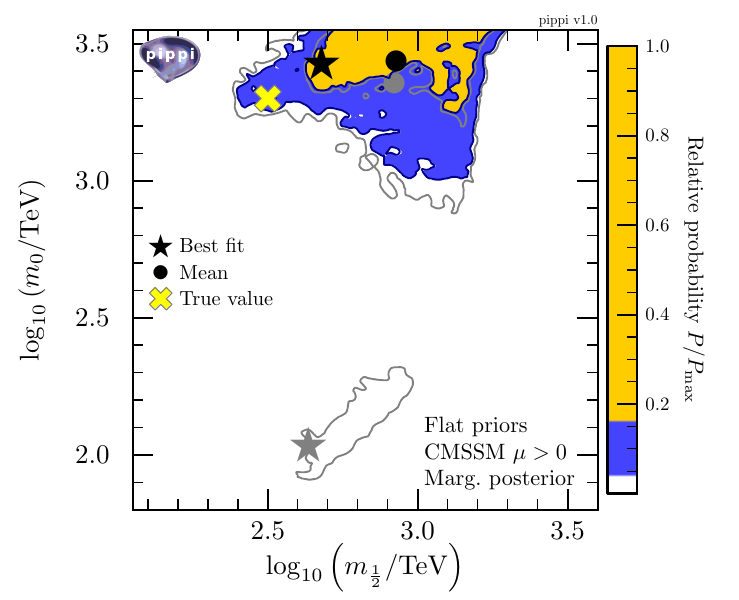}
\end{minipage}
\hspace{0.01\textwidth}
\begin{minipage}[b]{0.48\textwidth}
\centering
\includegraphics[height=0.85\linewidth]{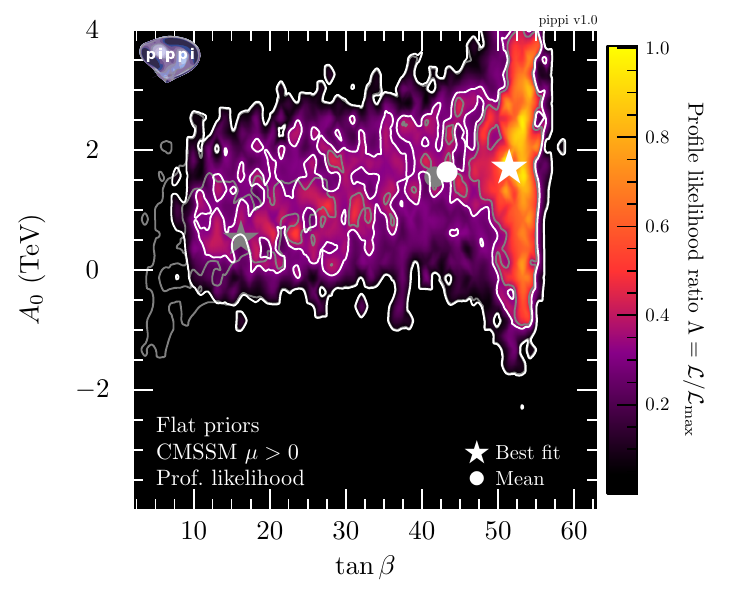}
\end{minipage}
\caption{Plots of posterior PDFs and profile likelihoods for a sample CMSSM chain taken from \protect\cite{Scott09c}.  All points with $m_0<1$\,TeV have been heavily down-weighted; grey lines in subplots a, c and d show the corresponding distribution / contours for the original chain, without down-weighting.  a) 1D marginalised posterior PDF of the parameter $m_{1/2}$, shown on a log scale.  Here a mock ``true point'' is plotted for illustration alongside the real best fit and posterior mean.  b) 2D posterior PDF of the parameters $A_0$ and $\tan\beta$ with 68\% and 95\% credible regions.  The corresponding profile likelihood contours are shown in grey.  c)  The 2D marginalised posterior PDF of the parameters $m_0$ and $m_{1/2}$, showing the corresponding posterior PDF contours from the original chain in grey.  A mock ``true value'' is again plotted.  d) 2D profile likelihoods for the parameters $A_0$ and $\tan\beta$, comparing the 68\% and 95\% confidence regions obtained in the original and down-weighted chains.}
\label{fig1}
\end{figure*}

\section{Functions}

\textsf{Pippi} consists of 6 core functions.  Options are specified via an ASCII \texttt{.pip} input file passed as a command-line argument.

\textsf{\textbf{pippi merge}} simply reads two or more chains, checks them for basic compatibility (number of columns, data types, etc.), and outputs a single concatenated chain to stdout.

\textsf{\textbf{pippi pare}} post-processes a chain, using a user-supplied function $F(\theta)$ for operating on a single sample $\theta$.  Pippi will dynamically load a python module $M$ whose name is passed as a command-line argument, find $F$ within it, and use $F(\theta)$ to operate on each point $\theta$ in the chosen chain.  It will then output the resulting post-processed chain to stdout.  The only thing required of the user is to write $F$, which implements the actual desired physics.  $F$ takes as input a vector containing the parameter and observable values of a single sample, and returns the post-processed parameter and observable values for that sample.  The returned sample (and hence the final post-processed chain) need not contain the same parameters and observables as the original chain, nor even the same number of them.  $M$ may contain any number of other routines, which may e.g.~open a data file and initialise a new likelihood component or observable to be added to the chain.  As of \textsf{pippi 2.0}, post-processing is also available via \textsf{pippi parse}.

\textsf{\textbf{pippi probe}} prints the names of the available data records in an hdf5 point database, along with the generalised column index that each will be mapped to.

\textsf{\textbf{pippi parse}} automatically bins a chain, then profiles its likelihood and/or marginalises its posterior PDF over relevant parameters.  Options include which parameters or combinations of parameters to profile/marginalise over, by how much each parameter or observable should be rescaled, whether it must be binned and displayed in terms of its actual value or logarithm, and the range of values that the bins should encompass.  The number of bins into which samples are sorted is individually configurable for each parameter or observable.  The final resolution with which bin centres will be interpolated between in the output data files is also configurable.  Either linear interpolation or curvature-minimising splines can be employed for this.  Unlike other parsing programs, the option to smooth the output distributions is explicitly excluded, as this amounts to modifying the underlying chain; a similar effect can be achieved whilst preserving the underlying data using interpolation.  \textsf{Parse} has the ability to work with an essentially arbitrary chain format, with multiplicities and likelihoods, $\chi^2$ values or $\pm$log-likelihoods located in any column of the chain.

Similarly, it can also import data from point databases in hdf5 format.  In this case, the user typically needs to specify in their input \texttt{.pip} file which named data records in the \texttt{.hdf5} file correspond to which parameters, observables, likelihoods, multiplicities and related quantities, in much the same way as different columns in ASCII chain files must be identified with specific quantities.  When importing results from \texttt{.hdf5} files, \textsf{pippi} operates in out of core mode, reading \textit{only} the observables and parameters requested for plotting into memory, rather than the entire file.  This essentially allows arbitrarily large datasets to be plotted; in extreme cases, this can be achieved by restricting single \texttt{.pip} files to deal with only one observable or pair of observables.  As of \textsf{pippi 2.0}, \textsf{parse} is also able to compute new observables and arbitrary transformations of old observables on the fly, by reading \textsf{Python} code embedded directly in a user's \texttt{.pip} file.  The results of these on-the-fly calculations can be saved and plotted like any other observable, and used in further calculations in the same run of \textsf{pippi}.  They can also be used together with the new ability to cut samples on the value of any parameter or observable to apply any arbitrarily complicated cuts to the underlying dataset before plotting.

\textsf{\textbf{pippi script}} writes shell scripts for plotting a parsed chain with \textsf{ctioga2}.  Either 1D or 2D distributions can be plotted, including comparison of two chains, or comparison of profile likelihoods and posterior PDFs.  1D plots may be presented as histograms or interpolated distributions.  2D plots may have arbitrary confidence contours, shading and a colour bar.  Axis labels and all other annotations can be specified directly in true \LaTeX.  The best fit and posterior mean may be plotted on or excluded from different plots, and corresponding legends and keys can be automatically drawn and placed.  A reference point (and key) can be specified and plotted in terms of any combination of parameters and/or observables.  A by-line can be placed in the top right of the figure, and a PDF logo or other image can even be included.  All aspects of the colour scheme, markers, gradients, transparencies and line drawing can be modified by choosing a different built-in scheme, or easily writing one's own scheme in a few short lines of \textsf{Python} code.

\textsf{\textbf{pippi plot}} runs the plotting scripts created in a \textsf{script} operation, and organises the resulting PDF files according to the specified pip file.

If \textsf{pippi} is invoked with only the name of a pip file, the \textsf{parse}, \textsf{script} and \textsf{plot} functions are automatically performed in this order.  Chains, intermediate and final files can all be arranged automatically into any combination of different or identical directories, using any combination of relative or absolute paths.  Missing paths are created automatically.

\section{Examples}

In Fig.~\ref{fig1} I show some example plots created from the chain included in the \textsf{pippi} distribution, which comes originally from \cite{Scott09c}.  This chain is based on a global fit to the Constrained Minimal Supersymmetric Standard Model (CMSSM), and was created using SuperBayeS v1.35 \cite{Trotta08}, with all likelihood components turned on.  Here I have used the \textsf{pare} function of \textsf{pippi} to reduce the likelihoods and posterior weightings of all points in the chain with values of the parameter $m_0<1\,$TeV, so as to remove the area at low $m_0$ known as the stau co-annihilation region.  The resulting 1D marginalised posterior PDF for the parameter $m_{1/2}$ is shown for the chain processed by \textsf{pippi pare} (a `pared chain') in blue in Fig.~\ref{fig1}a, alongside the corresponding marginalised posterior for the original chain in grey.  The equivalent 2D distribution in the $m_0$, $m_{1/2}$ plane is given directly below in Fig.~\ref{fig1}c, with the 68\% and 95\% credible contours from the pared chain plotted in colour, and the contours corresponding to the original chain in grey.  The best-fit and posterior mean are plotted in each case, in grey for the original and black for the pared chain.  For the sake of illustration, I have also added a fictional ``true value'' to the these two plots.

Fig.~\ref{fig1}b compares the posterior PDF (coloured) to the profile likelihood (grey) in the pared chain, this time in the $A_0$, $\tan\beta$ plane.  In this case I have employed a visual scheme with a gradient fill for the 2D marginalised posterior.  Similarly in Fig.~\ref{fig1}d, where I compare the profile likelihood in the pared (colour) and original (grey) chains in the $A_0$, $\tan\beta$ plane, using yet another built-in visual scheme.

An example pip file for creating these and other plots is included in the \textsf{pippi} distribution.  The \textsf{Python} function used with \textsf{pippi pare} to effect the $m_0>1\,$TeV post-processing cut is also included.  As of \textsf{pippi 2.0}, this example also includes the new features like data cuts and inline post-processing, so the final plots look a little different to Fig.~\ref{fig1}.

\section{Summary}
\label{summary}

\textsf{Pippi} can automatically bin, marginalise and profile sets of posterior or likelihood samples, or post-process them using functions easily defined by the user.  It produces clean, visually-appealing plots in native PDF format, with a minimum of effort and maximal flexibility.  \textsf{Pippi 2.0} depends on \textsf{Python} v2.7 or later, \textsf{ctioga2} v0.8 or later, \textsf{SciPy}, \textsf{NumPy} (v0.9.0 or later to use the spline interpolation option) and \textsf{bash}.  It requires essentially no installation beyond unpacking a tarball and adding the new directory to the shell PATH variable.  The latest incarnation of \textsf{pippi} can always be found at \href{http://github.com/patscott/pippi}{http://github.com/patscott/pippi}.

\section*{Acknowledgements}
I thank Antje Putze and Christoph Weniger for helpful comments during development of \textsf{pippi}, and Christoph for contributing draft code for supporting hdf5 inputs.

\bibliography{biblio,DMbiblio,SUSYbiblio,CosmoSF}

\begin{thebibliography}{10}
\providecommand{\url}[1]{\texttt{#1}}
\providecommand{\doi}[1]{\href{http://dx.doi.org/#1}{#1}}
\providecommand{\urlprefix}{URL }
\providecommand{\eprint}[2][]{\href{http://arxiv.org/abs/#2}{arXiv:#2}}

\bibitem{Allanach06}
B.~C. {Allanach}, C.~G. {Lester}, \prd{} \textbf{73}, 1, 015013 (2006),
  \eprint{hep-ph/0507283}, \doi{10.1103/PhysRevD.73.015013}

\bibitem{Trotta08}
R.~{Trotta}, F.~{Feroz}, M.~{Hobson}, et~al., \jhep{} \textbf{12}, 24 (2008),
  \eprint{0809.3792}, \doi{10.1088/1126-6708/2008/12/024}

\bibitem{Scott09c}
P.~{Scott}, J.~{Conrad}, J.~{Edsj{\"o}}, et~al., \jcap{} \textbf{1}, 31 (2010),
  \eprint{0909.3300}, \doi{10.1088/1475-7516/2010/01/031}

\bibitem{Akrami09}
Y.~{Akrami}, P.~{Scott}, J.~Edsj{\"o}, et~al., \jhep{} \textbf{4}, 57 (2010),
  \eprint{0910.3950}, \doi{10.1007/JHEP04(2010)057}

\bibitem{Akrami11DD}
Y.~{Akrami}, C.~{Savage}, P.~{Scott}, et~al., \jcap{} \textbf{4}, 12 (2011),
  \eprint{1011.4318}, \doi{10.1088/1475-7516/2011/04/012}

\bibitem{Akrami11Coverage}
Y.~{Akrami}, C.~{Savage}, P.~{Scott}, et~al., \jcap{} \textbf{7}, 2 (2011),
  \eprint{1011.4297}, \doi{10.1088/1475-7516/2011/07/002}

\bibitem{Fittino12}
P.~{Bechtle}, T.~{Bringmann}, K.~{Desch}, et~al., \jhep{} \textbf{6}, 98
  (2012), \eprint{1204.4199}, \doi{10.1007/JHEP06(2012)098}

\bibitem{CosmoMC}
A.~{Lewis}, S.~{Bridle}, \prd{} \textbf{66}, 10, 103511 (2002),
  \eprint{astro-ph/0205436}, \doi{10.1103/PhysRevD.66.103511}

\bibitem{ModeCode1}
M.~J. {Mortonson}, H.~V. {Peiris}, R.~{Easther}, \prd{} \textbf{83}, 4, 043505
  (2011), \eprint{1007.4205}, \doi{10.1103/PhysRevD.83.043505}

\bibitem{Putze10}
A.~{Putze}, L.~{Derome}, D.~{Maurin}, \aap{} \textbf{516}, A66 (2010),
  \eprint{1001.0551}, \doi{10.1051/0004-6361/201014010}

\bibitem{GalpBayes}
R.~{Trotta}, G.~{J{\'o}hannesson}, I.~V. {Moskalenko}, et~al., \apj{}
  \textbf{729}, 106 (2011), \eprint{1011.0037},
  \doi{10.1088/0004-637X/729/2/106}

\bibitem{NR3}
W.~H. Press, S.~A. Tuekolsky, W.~T. Vetterling, et~al., \emph{Numerical
  Recipes}, 3rd edn.{} (Cambridge University Press, 2007)

\bibitem{Skilling04}
J.~{Skilling}, in R.~{Fischer}, R.~{Preuss}, U.~V. {Toussaint}, eds.,
  \emph{American Institute of Physics Conference Series}, vol. 735, 395--405{}
  (2004)

\bibitem{MultiNest}
F.~{Feroz}, M.~P. {Hobson}, M.~{Bridges}, \mnras{} \textbf{398}, 1601 (2009),
  \eprint{0809.3437}, \doi{10.1111/j.1365-2966.2009.14548.x}

\bibitem{PIKAIA}
P.~{Charbonneau}, \apjs{} \textbf{101}, 309 (1995), \doi{10.1086/192242}

\bibitem{Storn}
R.~Storn, K.~Price, J.\ Global Optimization{} \textbf{11}, 4, 341 (1997),
  \doi{10.1023/A:1008202821328}

\bibitem{IC22Methods}
P.~{Scott}, C.~{Savage}, J.~{Edsj{\"o}}, et~al., \jcap{} \textbf{11}, 57
  (2012), \eprint{1207.0810}, \doi{10.1088/1475-7516/2012/11/057}

\bibitem{RooStats}
L.~Moneta, K.~Belasco, K.~Cranmer, et~al., PoS(\textbf{ACAT2010})057{}  (2010),
  \eprint{1009.1003}

\bibitem{barrett}
S.~{Liem}{}  (2016), \eprint{1608.00990}

\bibitem{superplot}
A.~{Fowlie}, M.~H. {Bardsley}, \epjp{} \textbf{131}, 391 (2016),
  \eprint{1603.00555}, \doi{10.1140/epjp/i2016-16391-0}

\end{thebibliography}

\end{document}